\documentclass[fleqn,twoside]{article}
 \usepackage{espcrc2}
\usepackage{amsmath}

\usepackage{graphicx}
\usepackage[figuresright]{rotating}

\newcommand{\Leff}{{\mathcal L}_{\rm eff}}
\newcommand{\LQCDQED}{{\mathcal L}_{{\rm QCD} \times {\rm QED}}(u, d, s,c, b)}
\newcommand{\nn}{\nonumber \\}

\title{Developments in the evaluation of $m_c$ dependent matrix
elements in $\overline B \to X_s \gamma$ at NNLO}

\author{Thomas Schutzmeier\address{Institut f\"ur Theoretische Physik
        und Astrophysik, Universit\"at W\"urzburg,\\
        Am Hubland, D-97074 W\"urzburg, Germany}}
       
\runtitle{Developments in the evaluation of $m_c$ dependent matrix
elements in $\overline B \to X_s \gamma$ }

\begin{document}

\begin{abstract}
The full prediction of the inclusive radiative $\overline B\to X_s\gamma$ decay rate at NNLO requires a complete evaluation of missing charm quark mass
dependent matrix elements. Recent developments in the on-going computations 
are reported and the current status is briefly overviewed.

\vspace{1pc}
\end{abstract}

\maketitle
\thispagestyle{empty}
\pagestyle{empty}

\section{Introduction}

One interesting candidate process in the indirect search for
non-standard physics is the rare inclusive $\overline B\to X_s\gamma$
decay \cite{Haisch:2007ic}.
Due to its low sensitivity to non-perturbative effects and,
being a flavour-changing neutral current, its loop-suppression in the 
Standard Model (SM)  stringent constraints on the parameter space of
physics beyond the SM can be derived from both accurate measurements
and precise theory predictions.

The latest measurements by BaBar, Belle and CLEO \cite{ref:exp}
have been combined by the Heavy Flavour Averaging Group (HFAG)
\cite{Barberio:2007cr} into the current world average (WA)
for the branching ratio,
\begin{equation}
  \mathcal B^{exp}_{E_\gamma > 1.9 GeV}
  = (3.15 \pm 0.23)\times 10^{-4}
\end{equation}
\normalsize
\noindent where a cut $E_{\gamma,0} > 1.6$~GeV has been imposed on the photon 
energy in the $\overline B$-meson rest frame. The first uncertainty
corresponds to a combined statistical and systematical error, the
second one is due to the theory input in the extrapolation of the
measured branching ratio to the reference value $E_{\gamma,0}$,
whereas the third one is connected to the subtraction of $b\to d\gamma$
contamination. On the theory side, the expected size of 
next-to-next-to leading order (NNLO) QCD effects to the partonic decay 
$b\to X_s^{partonic} \gamma$ is comparable with the overall
experimental error of about 7\% and therefore a complete SM
calculation at this level of accuracy is clearly needed.

The recent theoretical estimate of the branching ratio at the NNLO level
\begin{equation}
  \mathcal B^{exp}_{E_\gamma > 1.9 GeV}
  = (3.15 \pm 0.23)\times 10^{-4}
\end{equation}
was derived in \cite{Misiak:2006zs} after a large part 
of the NNLO program has been finished and
is in good agreement with the WA. The uncertainty here consists of
four types of error added in quadrature: non-perturbative (5\%),
parametric (3\%), higher-order (3\%) and $m_c$-interpolation ambiguity
(3\%).

Large logarithms of the form $\alpha_s(m_b)^n \log^m(m_b/m_W)$ appear
in QCD corrections to the partonic decay width $\Gamma(b\to s\gamma)$
and have to be resummed with renormalization-group techniques to get
a reasonable prediction.
Most suitably, this is done in the framework of an effective low-energy theory
with five active quarks by integrating out the heavy electroweak and the
top fields in the SM. As a consequence, local flavour-changing operators 
$Q_i(\mu)$ up to dimension six and Wilson coefficients $C_i(\mu)$ appear in the resulting effective Lagrangian $\Leff$.

A few years ago, the next-to-leading order QCD corrections to the $b\to s
\gamma$ decay have been completed (see e.g. \cite{Buras:2002er,Hurth:2003vb} 
and references therein).
The next-to-next-to leading order evaluation, which is a very
complicated task, is currently under way and large parts are already
finished. In general, three steps are required for a consistent calculation 
in the low-energy effective theory and in particular also 
at the NNLO level:
\begin{enumerate}
\item  Determination of Wilson
coefficients $C_i(\mu_0)$ at the electroweak scale $\mu_0=M_W$ by
requiring equality of Green's functions in the effective and full
theory at leading order in (external momenta)/$M_W$. To this
precision, the matching of the four quark operators $Q_1,\dots,Q_6$
and the dipole operators $Q_7, Q_8$ at the two-
and three-loop level, respectively, has been computed in 
\cite{Bobeth:1999mk,Misiak:2004ew}.
\item Derivation of the effective theory 
Renormalization Group Equations (RGE) and computation of the operator
mixing under renormalization by evolving the Wilson coefficients
$C_i(\mu)$ from $\mu_0$ down to the low scale $\mu_b \sim m_b$ using
the anomalous dimension matrix up to $\mathcal O(\alpha_s^3)$. In the
sectors $\{Q_1,\dots,Q_6\}$ and $\{Q_7,Q_8\}$, the three-loop
renormalization was found in \cite{Gorbahn:2004my,Gorbahn:2005sa}. 
Results for the four-loop mixing of
$\{Q_1,\dots,Q_6\}$ into $\{Q_7,Q_8\}$ were recently found by completing the
anomalous dimension matrix \cite{Czakon:2006ss}.
\item Calculation of on-shell matrix elements with single insertions of
effective operators at $\mu_b \sim m_b$ to $\mathcal O(\alpha_s^2)$.
This task is not completed yet, although a number of contributions is
known. In \cite{Melnikov:2005bx,Blokland:2005uk}, 
the two-loop matrix element of the dipole
operator $Q_7$ together with the corresponding bremsstrahlung
was determined, confirmed in \cite{Asatrian:2006ph} and subsequently extended to
include the full charm quark mass dependence in \cite{Asatrian:2006rq}.
Dominant contributions to the photon energy spectrum in the so-called large-$\beta_0$ approximation $\mathcal O(\beta_0 \alpha_s^2)$ have been obtained 
in \cite{Ligeti:1999ea}.
Using an expansion in $m_c^2/m_b^2$ the $\mathcal O(\beta_0
\alpha_s^2)$ contributions to the two-loop matrix elements of $Q_7$
and $Q_8$, as well as to the three-loop matrix elements $Q_1$ and
$Q_2$, were found in \cite{Bieri:2003ue}. 
These results have been confirmed in \cite{Boughezal:2007ny} and, 
moreover, the full fermionic 
corrections beyond the large-$\beta_0$ approximation have been provided there.
\end{enumerate}

The full matrix elements of $Q_1$ and $Q_2$ at $\mathcal
O(\alpha_s^2)$ constitute an important piece that is still missing. 
At NLO the choice of scale and scheme of $m_c$ constitutes 
the main source of uncertainty stemming from the fact, that these
operators contain the charm quark and contribute for the first time at
$\mathcal O(\alpha_s)$. Removing this ambiguity is therefore a NNLO
effect in the branching ratio. So far, the full matrix elements of the $Q_1$ and
$Q_2$ operators have been evaluated in the large $m_c$ limit, 
$m_c \gg m_b$, and subsequently used for an
interpolation to the physical range of $m_c$ \cite{Misiak:2006ab} 
assuming some ad-hoc value at $m_c = 0$. 
This is the source of the aforementioned interpolation ambiguity in the current NNLO branching ratio estimate. Removing this uncertainty requires the calculation of $\langle s\gamma|Q_{1,2}|b\rangle$ at physical $m_c$ and involves the evaluation of hundreds of on-shell three-loop vertex diagrams with two scales $m_b$ and $m_c$. Reducing this interpolation uncertainty is possible by computing the matrix elements at $m_c=0$ and thus fixing the endpoint of the interpolation. Both evaluations are currently being pursued. This paper is meant to describe the current status.

\section{The matrix elements $\langle s\gamma|Q_{1,2}|b\rangle$ at NNLO}

The effective low-energy theory, in which the calculation of 
$\langle s\gamma|Q_{1,2}|b\rangle $ at $\mathcal O(\alpha_s^2)$ is performed,
is given by the effective Lagrangian
\begin{eqnarray}\label{eq:effectivelagrangian}
\Leff&=& \LQCDQED \nonumber\\
&&+ \frac{4G_F}{\sqrt{2}} V^\ast_{ts} V_{tb} 
\sum^{8}_{i= 1} C_i (\mu) \, Q_i (\mu).
\end{eqnarray}
The first term corresponds to the usual QED-QCD Lagrangian for the
light SM fields, the second term gives the local operator product
expansion, $V_{ij}$ are the matrix elements of the
Cabibbo-Kobayashi-Maskawa matrix and $G_F$ is the Fermi coupling
constant.
The relevant physical operators are chosen as \cite{Chetyrkin:1996vx}
\begin{eqnarray} \label{eq:physicaloperators}
Q_{1,2} \, & = & \, (\bar{s} \Gamma_i c) (\bar{c} \Gamma'_i b), \\
Q_{3,4,5,6} \, & = & \, (\bar{s} \Gamma_i b) \sum\nolimits_q (\bar{q} \Gamma'_i
q) ,\nonumber
\end{eqnarray} \begin{eqnarray}
Q_7 \, & = & \, \frac{e}{16 \pi^2}\,{\overline m_b}(\mu)\, (\bar{s}_L 
  \sigma^{\mu \nu} b_R) \, F_{\mu \nu} , \nn
Q_8 \, & = & \, \frac{g}{16 \pi^2}\,{\overline m_b}(\mu)\, (\bar{s}_L 
  \sigma^{\mu \nu}T^a b_R) \, G^a_{\mu \nu} \nonumber.
\end{eqnarray}
where $\Gamma$ and $\Gamma'$ stand for various products of Dirac and
colour matrices.

As already mentioned, removing the interpolation uncertainty requires
the calculation of $Q_{1,2}$ matrix elements at the NNLO level. 
All appearing vertex diagrams have been generated, expressed through
scalar diagrams that depend on the two mass scales 
$m_c$ and $m_b$ and finally reduced with
Laporta's algorithm to 476 master integrals. The latter can be evaluated 
using two different approaches: numerical solutions of differential equations or the Mellin-Barnes technique. The first method consists of writing down differential equations for the master integrals in their kinematical invariants and solving them numerically after an expansion about some limit has been taken to provide the startpoint for integration with high precision. This was already used in the calculation of double fermionic QCD corrections to the photon polarization function \cite{Czakon:2007qi}. Moreover, in \cite{Czakon:2008zk} 
this idea was also applied to determine the full mass dependence of
the $t\overline t$-production cross-section 
from light quarks at NNLO. The second approach is based on the Mellin-Barnes 
(MB) technique, where MB representations can be derived in an
automatized way, analytically continued in $\epsilon$ and numerically
integrated utilizing the package \texttt{MB} \cite{Czakon:2005rk}. 
Both methods have been used to obtain
the full fermionic corrections, extending the so far known massless case 
to include heavy $b$ and $c$ quark loops \cite{Boughezal:2007ny}. 
As a result, it turned
out that while the charm quark contribution is well reproduced by 
the massless approximation, bottom quark loop insertions are overestimated
by a large extent. For the remaining bosonic parts the IBP reduction 
is still under way.
\begin{figure}\label{fig:O2O7graph}
    \begin{center}
        \includegraphics[width=0.25\textwidth]{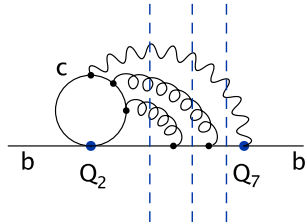}
    \end{center}
        \caption{Example graph}

\vspace*{-1cm}
\end{figure}
In the limit $m_c=0$ valuable information can be obtained from
$\langle s\gamma|Q_{1,2}|b\rangle$ to fix the endpoint of the
interpolation performed in \cite{Misiak:2006ab}. A possible way of getting $\langle s\gamma|Q_{1,2}|b\rangle$ is by interfering the matrix elements of the 
operators $Q_{1,2}$ with those of the dipole operator $Q_7$  and cutting 
the resulting four-loop propagators
selectively. The final state is required to contain at least one $s$ quark and 
one photon. 
Altogether, several hundreds of four-loop propagator-like diagrams are generated
with up to five-particle cuts and subsequently reduced to about 200
master integrals that have to be evaluated. The massless cases
among them have already been computed up to four-particle cuts using a
Mellin-Barnes based method. For this task, Mellin-Barnes representations
have been derived and numerically evaluated after performing the necessary 
phase space integrations. However, integrals containing massive lines 
from virtual $b$-quarks require a different approach. 
The method we are currently 
using in this respect is again based on differential equations (DEQ), but it 
turns out to be more involved as compared to the virtual corrections at $m_c
\neq 0$. As starting point for the expansion and integration of the system of 
DEQ we choose the large mass limit. Thus, boundary conditions are given by cut
integrals in the large mass limit and can be evaluated with 
automatized diagrammatic expansions and phase space integration. 
Unfortunately, the usual approach of numerical
integration up to the interesting kinematical point, to the on-shell 
limit in our case, suffers from divergences in the differential equations at 
both endpoints of the integration contour. 
This can be overcome by solving the system of DEQ with an expansion
ansatz around the on-shell point. Comparison of the numerical integration
at some point close to the on-shell kinematics with the expansion
fixes the boundary conditions of the latter. In the case of  
logarithmic on-shell divergences, the integrals are regularized with a
sufficient number of irreducible numerators and a change of the 
basis of master integrals.

The achievable accuracy of this method is, on the one hand,
given by the depth of the expansions at the starting and ending points on 
the integration contour and, on the other,
by the error pile-up during numerical 
integration. 
\begin{figure}[h]\label{fig:plots}
    \begin{center}
        \includegraphics[width=.2\textwidth]{PR90.epsi}\\[6mm]
        \includegraphics[width=.4\textwidth]{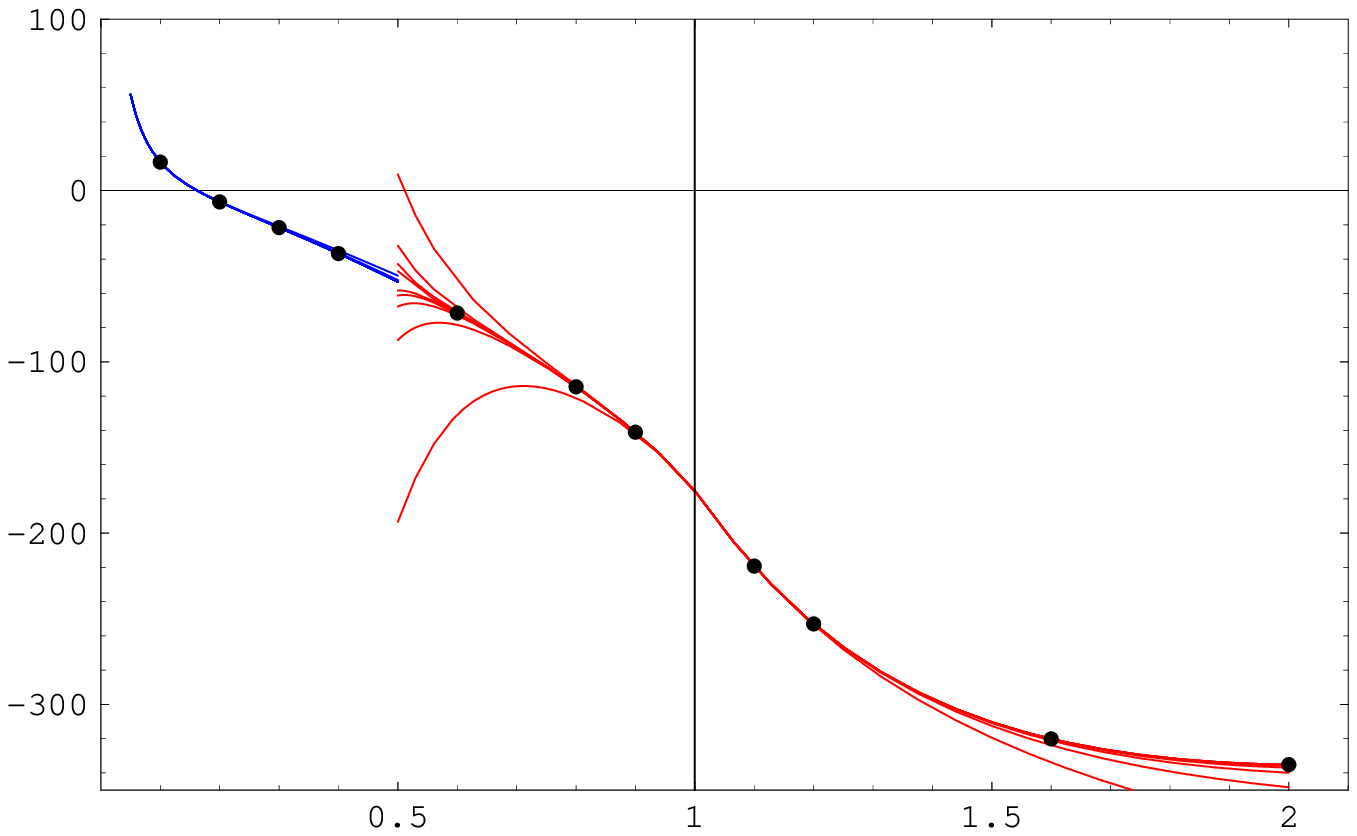}\\
        \footnotesize{$z=p^2/m^2$}
    \end{center}
    \caption{Example integral together with its plot as function of the ratio
    $z=p^2/m^2$. Dots denote the exact numerical results and the lines 
    at $z<0.5$ correspond to the series expansion for $z \to 0$. Curves
    at $z>0.5$ show different depths in the expansion around the
    on-shell point, the matching is performed in $z=0.9$}.

\vspace*{-1cm}
\end{figure}
Performing the numerical part with quadruple precision, 
a relative error of the order of $10^{-15}$ at the on-shell point seems 
to be obtainable.

As a first attempt we study diagrams involving two- and three-particle
cuts, which constitute the major part of the evaluation. Fig. \ref{fig:plots}
shows a plot of the $\mathcal O(\epsilon^0)$ term for one example integral. 
The expansion in the large mass limit has been performed up to 20 
terms in $z=p^2/m_b^2$ whereas in the on-shell 
limit 12 terms have been taken into account. The numerical integration
is performed starting at $z=0.05$ and the matching of the numerics and
the on-shell expansion is done at $z=0.9$. It is apparent, that both
series converge nicely against the numerical points. 

\vspace*{.5mm}
The first three terms of the expansion in the limit of $z = p^2/m^2 \to 0$ 
of the integral in Fig. \ref{fig:plots} are given by
{\footnotesize
\begin{eqnarray*}
&I_{z\to0}&= \nn
&&  \pi \left\{\frac{1}{\epsilon^3}\left[-\frac{2}{3}\right]  + 
   \frac{1}{\epsilon^2}\left[-\frac{10}{3} - \frac{1}{2} z - \frac{1}{9} z^2
- 2\log(z)\right] 
\right.\nn&&\left.
   + \frac{1}{\epsilon} \left[-\frac{34}{3} +
   z \left(-\frac{9}{4} - \frac{\zeta_2}{2}
   - \frac{3}{2} \log(z) \right) + 
\right.\right.\nn&&\left.\left.
   z^2 \left(-\frac{77}{54} + \frac{2}{9} \zeta_2 - 
       \frac{1}{3} \log(z)\right)\right]
\right.\nn&&\left.
 + \left[
-\frac{98}{3} + 4 \zeta_2 - \frac{28}{9}\zeta_3 - 34 \log(z) - 15 \log(z)^2 
\right.\right.\nn&&\left.\left.
- 3 \log(z)^3 + z\left(-\frac{51}{8} - \frac{21}{4} \zeta_2 + \frac{5}{2} \zeta_3 + 
\right.\right.\right.\nn&&\left.\left.\left.
\left(-\frac{27}{4} - \frac{3}{2} \zeta_2\right) \log(z) - \frac{9}{4} \log(z)^2\right) + 
\right.\right.\nn&&\left.\left.
 z^2 \left(-\frac{2047}{324} + \frac{13}{9} \zeta_2 - \frac{10}{9} \zeta_3 +
\right.\right.\right.\nn&&\left.\left.\left.
 \left(-\frac{77}{18} + \frac{2}{3}\zeta_2\right)\log(z) 
 - \frac{1}{2}\log(z)^2\right)
 \right]
\right\}.
\end{eqnarray*}
}\normalsize
The corresponding first few terms of the expansion around the 
on-shell kinematical point with $y = z^{-1} - 1$ read
{\footnotesize
\begin{eqnarray*}
&I_{z\to1}&= \nn
&& -2.0944 \frac{1}{\epsilon^3}
+ \frac{1}{\epsilon^2}\left(-12.703 + 10.335\, y - 
\right.\nn && \left.
    4.7124\, y^2\right) + \frac{1}{\epsilon}\left(
    -52.607 + 81.505\, y - 67.338\, y^2\right)
\nn && 
+ \left(-175.32 + 454.59\, y - 472.67\, y^2\right).
\end{eqnarray*}
}
\normalsize
\section{Conclusions}
An evaluation of the missing matrix elements $\langle
s\gamma|Q_{1,2}|b\rangle$ at $\mathcal O(\alpha_s^2)$ is essential for
the reduction of the current uncertainty in the estimate of the branching ratio 
$\mathcal B(\overline B \to X_s\gamma)$. Two different approaches are
used and will eventually reduce or even remove the 
remaining $m_c$-interpolation ambiguity that amounts to about 3\%.

\section{Acknowledgements}
This work is supported by the Sofia Kovalevskaja Award of the
Alexander von Humbold Foundation sponsored by the German Federal
Ministry of Education and Research.

\end{document}